\documentstyle[aps,prl,twocolumn]{revtex}
\begin{document}

\noindent {\bf Balatsky and Salkola Reply}:
In their Comment, Aristov and Yashenkin \cite{AY} claim that 
we \cite{BS} have calculated the impurity-induced wave function
incorrectly and hence missed a power-law term that decays as $1/r$
to all directions. On the contrary to their claim, we have explicitly 
noted the presence of this term in the impurity wave function as is evident
from the discussion following Eq.~(1) in \cite{BS}. Moreover, we have explored
the spatial dependence of the impurity state in \cite{SBS1} where
all the terms were kept (see also Ref.~\cite{SBS2}). That there is a
$1/r$ term to all directions is a simple observation which follows,
for instance, from the calculation with the linearized quasiparticle 
spectrum near the nodes. Nonetheless, we decided to neglect it in our
subsequent discussion of localization for the following reasons.

First, if we take into account the $1/r$ decay of the wave function for all
directions, the impurity state is not normalizable. The fact that the 
single-particle states are already extended makes the whole issue of 
localization more subtle. However, we did not address this question, because
the hybridization of impurity states with the quasiparticle continuum
regularizes the problem naturally by introducing a cut-off at large 
distances.

Second, for $\Delta_0 \ll E_F$, the numerical solution of the impurity 
state clearly shows the strong cross-shaped anisotropy. Indeed, with good 
approximation the impurity wave function is concentrated in the cross-shaped 
tails as shown in Fig.~2 of Ref.~\cite{SBS1} or in Fig.~8 of Ref.~\cite{SBS2}.

Third, at nonzero impurity concentrations, the life-time effects cut off the
power-law behavior at finite lengths. As an illustration, consider
a $d$-wave superconductor close to half filling so that the Fermi 
surface is square. Let $\bf n$ be the normal to the two sheets of the
Fermi surface which are most nearly normal to ${\bf r}$, and 
define
$r_\perp =|{\bf n}\cdot {\bf r}|$ and $r_\parallel =|({\bf e}_3\times{\bf n})\cdot {\bf r}|$; by definition, $r_\perp \ge r_\parallel$. Then
\begin{eqnarray}
& & G_3({\bf r},\omega\rightarrow 0)  \ \propto \\
& & {(r_\perp/\xi_\perp)\sin k_{F\perp}r_\perp \over
\sqrt{(r_\perp/\xi_\perp)^2 + (r_\parallel/a)^2}}
K_1\Big(\sqrt{(r_\perp/\lambda_\perp)^2 + (r_\parallel/\lambda_\parallel)^2}\Big),\nonumber
\end{eqnarray}
where $K_1$ is the modified Bessel function of the first order, 
$\xi_\perp = a(W/\Delta_0)$ is the coherence length,
$\lambda_\perp = \xi_\perp(\Delta_0/\gamma)$
and $\lambda_\parallel = a(\Delta_0/\gamma)$ are the cut-off length scales. 
Here, $\gamma$ is the inverse life time, $W$ is the half bandwidth, and $a$ is 
the lattice spacing. Note that, for $\Delta_0 \ll W$, 
the anisotropy of the impurity state is fully developed 
because $\xi_\perp \gg a$. Furthermore, the decay rate along 
the nodal directions is much slower than in the other directions: 
$\lambda_\perp \gg \lambda_\parallel$. For $r_\perp < \lambda_\perp$
 and $r_\parallel < \lambda_\parallel$, we obtain the usual power-law 
behavior
and, for $r_\perp > \lambda_\perp$ or $r_\parallel > \lambda_\parallel$, 
the impurity state is exponentially small to all directions
due to the factors $e^{-r_\perp/\lambda_\perp}$ and
$e^{-r_\parallel/\lambda_\parallel}$. Thus, we may conclude that the 
power-law decay 
for all but the nodal directions is less important. As a consequence, 
the original problem of long-range
hoppings with the exponential cut-off is mapped in the limit of a dilute
 concentration of unitary impurities onto an effective
tight-binding model with the hopping range $\lambda_\perp \gg \xi_\perp$.
In this model, weak localization leads to an extremely long localization
length compared to the bare one in the absence of the impurity-induced
quasiparticle states.

Finally, Aristov and Yashenkin argue that all the physical 
consequences of our picture of highly anisotropic impurity states 
``appear to be incorrect''. Nevertheless, they arrive at 
the conclusion similar of \cite{BS} regarding the absence of
localization. It is  clear that, along the directions $r_\perp \sim r_\parallel$, 
the $(\Delta_0/W) (a/r_\parallel)$ term will be unimportant
at small values of $\Delta_0/W$. In over-doped high-temperature
superconductors, this ratio is small, whereas in the 
optimally doped regime it is on the order of 0.2. 

We conclude by noting that long-range hopping interactions require a dilute
concentration of unitary impurities. At high impurity concentrations, 
the hopping interactions are of short range and less anisotropic. 
There are two sources of anisotropy of the impurity state: one is 
the anisotropy of the energy gap and another is the anisotropy of 
the Fermi velocity, which can become important close to half filling 
and at a high impurity concentration.

We would like to thank G.E.Volovik and X.G.Wen for earlier discussions.

\

\noindent
A.V.Balatsky$^1$ and M.I.Salkola$^2$\hfill \today 

\

\noindent
$^1$Theoretical Division\\
Los Alamos National Laboratory\\
Los Alamos, New Mexico 87545\\

\noindent
$^2$Department of Physics\\
Stanford University\\
Stanford, California 94305

\end{document}